# SILICON SHEETS BY REDOX ASSISTED CHEMICAL EXFOLIATION

Mohamed Rachid Tchalala[1,2], Mustapha Ait Ali[2*], Hanna Enriquez[1], Abdelkader Kara[3*], Abdessadek Lachgar[4], Said Yagoubi[5], Eddy Foy[6], Enrique Vega[6], Azzedine Bendounan[7], Mathieu G. Silly[7], Fausto Sirotti[7], Serge Nitshe[8], Damien Chaudanson[8], Haik Jamgotchian[8], Bernard Aufray[8], Andrew J. Mayne[1], Gérald Dujardin[1] and Hamid Oughaddou[1,9*]

[1] Institut des Sciences Moléculaires d'Orsay, ISMO-CNRS, Bât. 210, Université Paris-Sud, F-91405 Orsay, France

[2] Laboratoire de Chimie de Coordination et Catalyse, Département de Chimie, Faculté des Sciences-Semlalia, Université Cadi Ayyad, Marrakech, 40001, Morocco

[3] Department of Physics, University of Central Florida, Orlando, FL 32816, USA

[4] Department of Chemistry, Center for Energy, Environment and Sustainability, Wake Forest University, , Winston Salem, NC 27109 USA

[5] LEEL SIS2M UMR 3299 CEA-CNRS-Université Paris-Sud 11, CEA Saclay, F-91191 Gif-Sur-Yvette, France

[6] LAPA SIS2M UMR3299 CEA-CNRS, CEA Saclay, 91191 Gif-sur Yvette, France

[7] TEMPO Beamline, Synchrotron Soleil, L'Orme des Merisiers Saint-Aubin, B.P. 48, F-91192 Gif-sur-Yvette Cedex, France

[8] Aix-Marseille Université, CNRS, CINaM UMR 7325, 13288 Marseille, France

[9] Département de Physique, Université de Cergy-Pontoise, F-95031 Cergy-Pontoise Cedex, France

[*] Corresponding authors:

abdelkader.kara@ucf.edu; hamid.oughaddou@u-psud.fr; aitali@uca.ma

**Abstract:**

In this paper, we report the direct chemical synthesis of silicon sheets in gram-scale quantities by chemical exfoliation of pre-processed calcium di-silicide ($CaSi_2$). We have used a combination of X-ray photoelectron spectroscopy, transmission electron microscopy and Energy-dispersive X-ray spectroscopy to characterize the obtained silicon sheets. We found that the clean and crystalline silicon sheets show a 2-dimensional hexagonal graphitic structure.

A substantial amount of recent advances in nanoscience has focused on two dimensional (2D) sheet-like structures. Silicene is considered to be a novel 2D material for nano-electronics as it will naturally benefit from the vast existing Si-based R&D infrastructure. For instance, the fabrication of the electrical contacts, which is an important problem for materials such as graphene, could be facilitated in the case of silicene by using silicides [1]. One of the main differences between carbon and silicon is the tendency for carbon to form $sp^2$ hybridization while silicon prefers $sp^3$ hybridization. This is reflected by the fact that silicon crystallizes in a diamond-like structure while graphitic carbon structure is the most stable form. Previous studies of one-dimensional (1D) silicon based systems by theoretical methods built silicon nanotubes from hexasilabenzenes with pure $sp^2$ hybridization [2-6]. However, experimental data points towards partial $sp^2$ hybridization in single walled Si-nanotubes [7-9]. It is reported that silicene presents a linear dispersion at high symmetry points of the Brillouin zone and consequently, its charge carriers behave as massless relativistic particles [10-13] Therefore, all the expectations associated with graphene, such as high-speed electronic nanometric devices based on ballistic transport at room temperature, could be transferred to this new silicon material with the advantage of being compatible with existing semiconductor devices. Silicene has been grown by epitaxy on silver [14-18], gold [19] and Irridium [20] surfaces under ultrahigh vacuum conditions (UHV). It is clear, however, that studying and understanding the properties of silicene, and the consequent development of silicene-based semiconducting devices, requires either to grow silicene on insulating materials, or to prepare bulk silicene by chemical methods.

In this paper we report the direct chemical synthesis of silicon sheets in gram-scale quantities by redox-assisted chemical exfoliation (RACE) of calcium disilicide ($CaSi_2$). The latter consists of alternating Ca layers and corrugated Si (111) planes as shown in Figure 1.

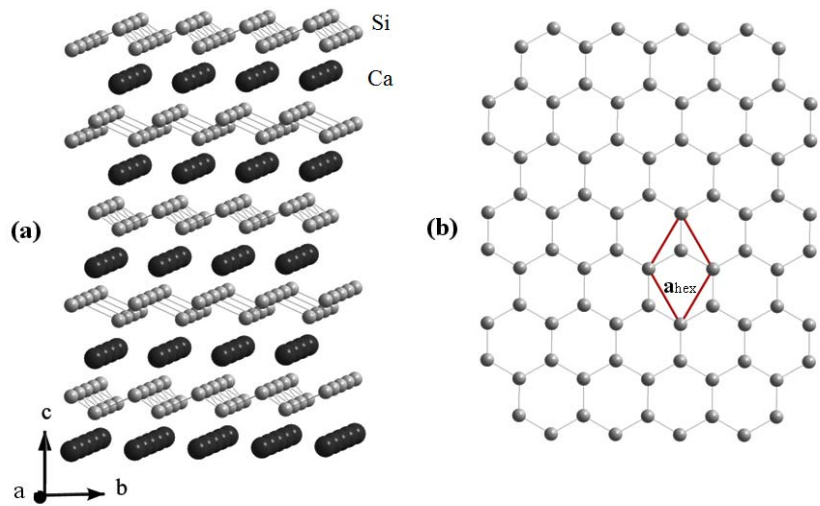

**Figure 1: (a)** Crystal structure of CaSi$_2$ (cell parameters: a=3.855 Å and c=30.60 Å with space group R-3m). The large circles represent Ca atoms and the small circles represent Si atoms. **(b)** Silicon sheet in CaSi$_2$ crystal structure.

Silicene with a honeycomb graphene-like structure has been observed recently by scanning tunneling microscopy (STM) on Ag(110) and on Ag(111) surfaces [14,15]. Silicene formation has also been studied on Au(110) [19], Ir(111) [20] and on ZrB$_2$ [21] surfaces . Furthermore, silicene NRs deposited on Ag(110) surface have shown reduced reactivity to oxygen compared to a Si(111) layer [22] which makes them promising candidates for stable devices. So far, these silicon structures have been grown using epitaxial growth techniques under UHV. However, chemical processes may provide an alternative route to large-scale synthesis of 2D silicon nanomaterials under production conditions. To the best of our knowledge, only a few investigations have been performed to synthetize 2D silicon structures [23]. From a chemical point of view, it is well known that silicon forms a variety of stable binary Zintl compounds [24] in which silicon forms the anionic part of the structure and alkali, alkaline earth or lanthanide metals form the cationic part [25-27]. Among the Zintl silicides, only calcium disilicide (CaSi$_2$) has a structure of interconnected $Si_6$ rings forming anionic 2D silicon puckered sheets $(Si_2)_\infty^{2-}$ held together by $Ca^{2+}$ that form planar monolayers (Figure 1). The covalent structure of the silicide layers is stable and remarkably flexible [28-30] and can be chemically modified by oxidation, leading to removal of the intercalated metal ions. Using these experimental observations, a synthesis of a 2D Si backbone of siloxene nanosheets $(Si_6H_3(OH)_3)$ was reported in an earlier publication [23]. Unfortunately the silicon skeleton was partially oxidized [31]. Nakano et al. reported the synthesis of Mg-doped silicon sheets - capped with oxygen - by exfoliation of Mg-doped $CaSi_2$ [32]. Despite these studies, the chemical synthesis of pure silicon sheets has not yet been reported. In the present study, we report results on the synthesis of silicon sheets by chemical exfoliation of $(Si_2)_\infty^{2-}$ from $CaSi_2$ using a modified procedure published by Nakano et al.[30] We coin this method redox-assisted chemical exfoliation (RACE). Let us recall that $CaSi_2$ is an ionic $(Ca^{2+}(Si_2)^{2-})$ material, and the reduction of the charge on the (negatively charged) silicon layers is an important step in order to diminish the strong electrostatic interaction between the $Ca^{2+}$ and $(Si_2)^{2-}$ layers [32].

A 150-ml Schlenk tube [33] with a magnetic stirring bar is charged with 2.66 mmol $CaSi_2$ (Aldrich, Ca: 30-33% and Si: 60-62%) (255.1 mg) and 5.32 mmol of metallic potassium (Merck, 96%) (208 mg). The tube is purged with $N_2$ three times and heated under a pressure of 0.01mbar at 160 °C for 5 hours. After cooling to room temperature (RT), a degassed solution of 10g isopropylamine hydrochloride (iPA HCl) in 50 mL of ethanol is gradually added under nitrogen until the amalgam is obtained. The mixture is degassed by three vacuum-filling nitrogen cycles and magnetically stirred at RT for 24 hours to form a green-brown precipitate of fine particles. Subsequently 50 mL of degassed water was added to the tube, and the mixture was allowed to decant for a period of 30 to 40 minutes to form a black deposit at the bottom of the Schlenk tube which was found to be unreacted $CaSi_2$ as confirmed by powder X-ray diffraction. The floating particles (less dense than water) are recovered by filtration and the solid is washed with cold water 3 times to remove any formed chloride salts. The green-brown solid is dried at RT under vacuum for 4 hours yielding silicon sheets corresponding to 20% of the used $CaSi_2$.

In the original work by Nakano *et al.* [32] Mg-doped $CaSi_2$, ($CaSi_{1.85}Mg_{0.15}$) was used as a precursor to the chemical synthesis of silicon sheets. They prepared $CaSi_{1.85}Mg_{0.15}$ by melting a mixture of CaSi, Si, and Mg ; then the material obtained was treated in a solution of propylamine hydrochloride (PA·HCl), resulting in the de-intercalation of $Ca^{2+}$ ions and the formation a light-brown suspension containing silicon nanosheets (SiNSs). The authors proposed the following mechanism for the exfoliation process: 1) oxidation of $CaSi_{1.85}Mg_{0.15}$ initiated by the oxidation of Ca atoms with PA·HCl, accompanied by the liberation of PA; 2) the presumably very reactive Mg-doped $Si_6H_6$ obtained is readily oxidized by water to form gaseous hydrogen; and 3) Mg-doped layered silicon with capping oxygen atoms is exfoliated by reaction with the aqueous solution of PA. This mechanism helped us to design and propose the following scheme for the present exfoliation. In our case the exfoliation occurs in the following steps.

(1) Diffusion of potassium (K) in $CaSi_2$ at 160 °C, leads to the formation of the $K_{2x}Ca_{1-x}Si_{2-x}$ phase and a small amount of KSi and Ca metal driven by the reduction of some $Ca^{2+}$ by K, as shown in the following chemical equation (1) :

$$CaSi_2 \;+\; 3xK \;\rightarrow\; K_{2x}Ca_{1-x}Si_{2-x} \;+\; xKSi \;+\; xCa \qquad (1)$$

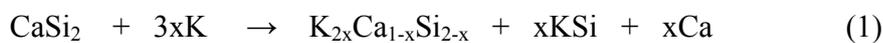

(2) The treatment of the mixture obtained (amalgam) by an ethanolic solution of isopropylamine hydrochloride (*i*PA·HCl) leads to the formation of Si nanosheets (SiNSs). These steps can be summarized by the following three chemical equations:

$$K_{2x}Ca_{1-x}Si_{2-x} + 2\ iPA.HCl \rightarrow 2xKCl + (1-x)CaCl_2 + (2-x)\ Si\ (SNSs) + 2iPA + H_2 \quad (2)$$

$$Ca + 2\ iPA.HCl \rightarrow CaCl_2 + 2iPA + H_2 \quad (3)$$

The excess potassium reacts with iPA.HCl to release $H_2$ as follows:

$$K + iPA.HCl \rightarrow KCl + 2iPA + 1/2H_2 \quad (4)$$

We have analyzed the synthesized sheets by X-ray photoemission spectroscopy (XPS), high-resolution transmission electron microscopy (HR-TEM) as well as energy dispersive X-ray spectroscopy (EDX). The Si 2p core level was probed by XPS at the SOLEIL Synchrotron Radiation facility on the TEMPO beam line, at binding energies (BE) of ~100 eV (Figure 2).

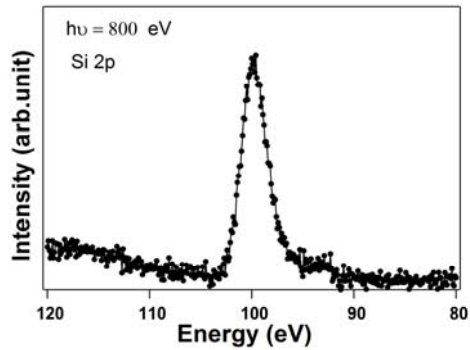

**Figure 2** : XPS spectrum of the synthesized Si sheets. The Si 2p core levels is shown

Note that within the precision of the instrument, the Si binding energies are very close to those of bulk silicon.[31] As a matter of fact, the XPS measurements show that the synthesized sheets contain almost exclusively Si atoms. In addition, to check the oxidation trace in the silicon sheets we have performed Energy-dispersive X-ray spectroscopy (EDX) experiments. Figure 3 shows different spectra corresponding to the analysis of different grains. Figure 3a corresponds to the starting $CaSi_2$ powder showing both Si and Ca peaks. Figures 3b and 3c correspond to grains of silicon sheets indicating that the Ca atoms have been completely removed. However in Figure 3b we observe a small trace of oxygen which is due to exposure of the powder to air. Let us note that the observed Al peak on all figures is due the sample holder which is made from Aluminum.

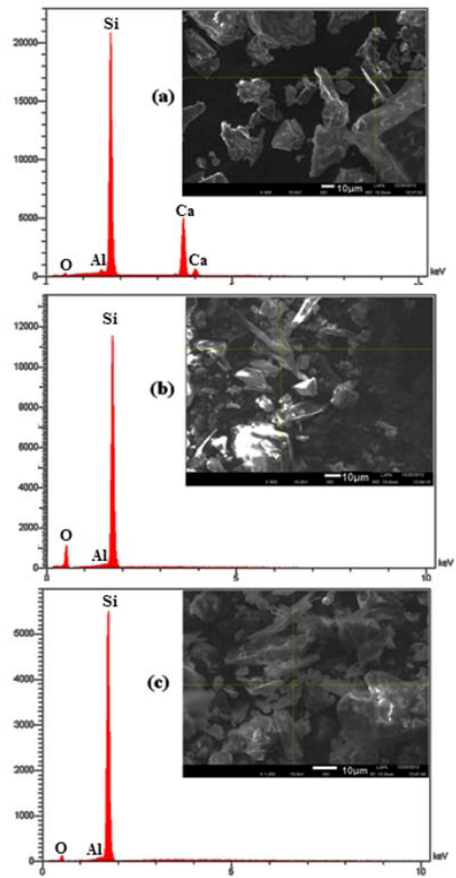

**Figure 3:** EDX spectra of selected area of grains: (a) starting material ($CaSi_2$); (b) and (c) of the prepared sample by the RACE method**.**

In order to provide further insight into the atomic structure of the prepared silicon sheets, we have performed HR-TEM measurements. Figure 4a reveals that the synthesized silicon product consists of multiple silicon sheets. At high resolution, the silicon sheets are highly ordered (Figure 4b). The diffraction pattern in Figure 4c shows a hexagonal spot arrangement. This provides clear evidence that the silicon sheets are well ordered and crystalline.

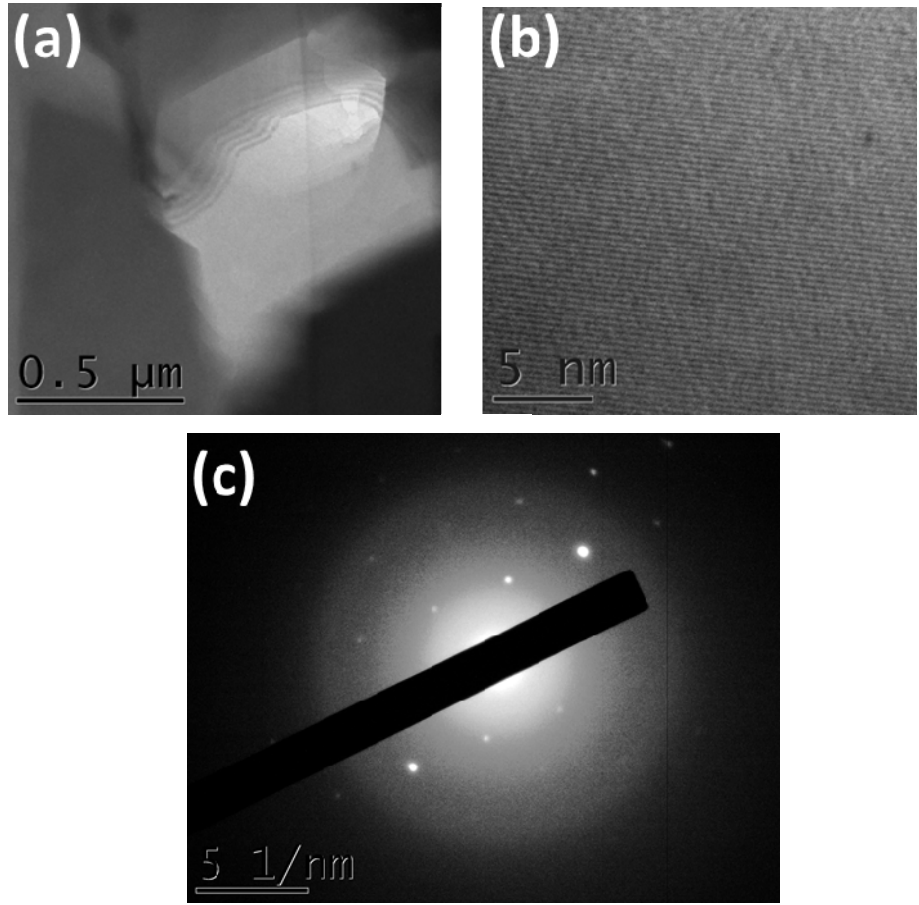

**Figure 4:** a) TEM image showing a stack of silicon sheets, b) TEM image showing a highly oriented crystalline silicon sheet, c) ED pattern recorded perpendicular to the surface of the sheet. This hexagonal pattern originates either from a quasi-two dimensional hexagonal lattice directly, like that of silicene, graphene and carbon nanotubes, or from a (111) oriented sheet (puckered layer) of a Si bulk diamond lattice. For the hexagonal system, the observed spots correspond to the four (100) and the two (110) reflections. The first spot of the diffraction pattern located at $3.49 \pm 0.20$ nm$^{-1}$ from the center gives an interlayer spacing ($d_{100}$ or $d_{110}$) of $0.286 \pm 0.016$ nm in the direct lattice. From the distance $d_{100}$ (or $d_{110}$), equal to $a_{hex} \times \sqrt{3}/2$ (where $a_{hex}$ is the 2D hexagonal lattice constant, see Figure 1b), we deduce a value of $a_{hex}$ equal to $0.330 \pm 0.018$ nm. This value is smaller than the 0.38 nm corresponding to the hexagonal lattice constant of the

(111) oriented layers of a silicon diamond lattice. Hence the structure of the synthetized Si sheets cannot correspond to the (111) oriented layer of Si diamond lattice. Therefore, these silicon sheets may correspond to a 2D honeycomb Si lattice stacked in planes with a surface lattice constant $a_{hex}$ = 0.330 ± 0.018 nm, corresponding to a distance between the first Si atoms neighbors of 0.19 nm. Let us recall that all the previous studies report that the silicene is corrugated [15,17,20,35,36-37] with a buckling between 0.04 and 0.12 nm [11,34], resulting in a projected distance (in the plane of the hexagons) as low as 0.19 nm. Hence, the first neighbor Si-Si distance in the present study is close to those reported for silicene sheets (0.20, 0.22, 0.25 nm) [15,17,35].

In conclusion, we have synthesized silicon sheets by a simple room temperature procedure called redox-assisted chemical exfoliation of calcium disilicide ($CaSi_2$). The data show clearly that the synthesized Si sheets do not correspond to the (111) oriented layer of Si diamond lattice, instead they have a hexagonal graphitic-like structure resembling that of silicene. In order to increase the size of sheets, an improved synthesis is under investigation using an appropriate stabilizer, which will prevent the sheets to aggregate.

**Acknowledgements**

R.T. and H.O. acknowledge the financial support from the European Community FP7-ITN Marie-Curie Programme (LASSIE project, Grant Agreement No.238258) and the help of Professor J. L. Lemaire its scientific coordinator in France at the Observatoire de Paris. Beam time was allocated under the SOLEIL Project No. 20120189. AK and MA would like to thank ISMO and STARM for support.